\documentclass[12pt]{article}
\usepackage{graphicx}
\catcode`\@=11
\def\numberbysection{\@addtoreset{equation}{section}
\def\theequation{\thesection.\arabic{equation}}}

\numberbysection
\newcommand{\be}{\[}
\newcommand{\bea}{\begin{eqnarray*}}
\newcommand{\beq}{\begin{equation}}
\newcommand{\beqa}{\begin{eqnarray}}
\newcommand{\ee}{\]}
\newcommand{\eea}{\end{eqnarray*}}
\newcommand{\eeq}{\end{equation}}
\newcommand{\eeqa}{\end{eqnarray}}
\newcommand{\abs}[1]{\vert#1\vert}
\newcommand{\ap}{_{\rm ap}}
\newcommand{\chat}{{\widehat\chi}}
\newcommand{\cond}{{\rm cond}}
\newcommand{\cu}[1]{\langle\!\langle#1\rangle\!\rangle}
\renewcommand{\d}{{\rm d}}

\newcommand{\dif}{{\rm dif}}
\newcommand{\dom}{{\rm dom}}
\newcommand{\e}{{\rm e}}
\newcommand{\ev}{{\rm even}}
\newcommand{\frad}[2]{\displaystyle{\displaystyle#1\over\displaystyle#2}}
\renewcommand{\i}{{\rm i}}
\newcommand{\lam}{\lambda}
\newcommand{\mean}[1]{\langle#1\rangle}
\newcommand{\od}{{\rm odd}}

\newcommand{\ret}{{\rm ret}}
\newcommand{\s}{\sigma}
\newcommand{\spin}{{\rm spin}}
\newcommand{\st}{{^\star}}
\newcommand{\toinf}{\mathrel{\mathop{\longrightarrow}\limits_{N\to\infty}}}
\newcommand{\tot}{\leftrightarrow}
\newcommand{\var}{\mathop{\rm Var}\nolimits}
\newcommand{\C}{{\cal C}}
\renewcommand{\H}{{\cal H}}
\newcommand{\N}{{\cal N}}
\renewcommand{\P}{{\cal P}}
\renewcommand{\S}{\Sigma}
\newcommand{\Sp}{{\cal S}}
\newcommand{\T}{{\cal T}}
\newcommand{\1}{{\bf 1}}

\begin{document}
\centerline{\Large\bf Metastable states of the Ising chain}
\vspace{.3cm}
\centerline{\Large\bf with Kawasaki dynamics}
\vspace{1.6cm}
\centerline{\large
G.~De Smedt$^{a,}$\footnote{desmedt@spht.saclay.cea.fr},
C.~Godr\`eche$^{b,}$\footnote{godreche@spec.saclay.cea.fr},
and J.M.~Luck$^{a,}$\footnote{luck@spht.saclay.cea.fr}}
\vspace{1cm}
\centerline{$^a$Service de Physique Th\'eorique\footnote{URA 2306 of CNRS},
CEA Saclay, 91191 Gif-sur-Yvette cedex, France}
\centerline{$^b$Service de Physique de l'\'Etat Condens\'e,
CEA Saclay, 91191 Gif-sur-Yvette cedex, France}
\vspace{1cm}

\begin{abstract}
We consider a ferromagnetic Ising chain
evolving under Kawasaki dynamics at zero temperature.
We investigate the statistics of the
metastable configurations in which the system gets blocked
(statistics of energy, spin correlations, distribution of domain sizes).
A systematic comparison is made with analytical predictions
for the ensemble of all blocked configurations
taken with equal a priori weights (Edwards approach).
\end{abstract}

\vfill
\noindent To be submitted for publication to the European Physical Journal B

\noindent P.A.C.S.: 05.70.Ln, 64.60.My, 75.40.Gb.

\newpage
\setcounter{footnote}{0}
\section{Introduction}

Glassy dynamics is often described as motion
in a complex energy landscape~\cite{gold},
with many valleys separated by barriers.
Valleys have been given various definitions~\cite{tap,ktw,sw,fv},
which are not equivalent from a dynamical point of view~\cite{bir}.
One common feature of all these approaches
is the exponential growth with the system size
of the number of valleys at fixed energy density $E$:
\beq
\N(N;E)\sim\exp(N S\ap(E)),
\label{ne}
\eeq
where $S\ap(E)$ is the `a priori' configurational entropy, or complexity.

As valleys are so numerous, an important issue
concerns the possibility of giving an a priori statistical description of them.
The key question to address is then the following.
Assuming the initial configuration be random,
does the system sample all the possible valleys with equal statistical weights,
i.e., with a uniform or flat measure, or, to the contrary,
is the size of the basin of attraction of each valley relevant?
The assumption that an a priori statistical description
with a flat measure could hold
was first formulated by Edwards~\cite{edwards}
in the context of granular materials, and is therefore
referred to as the Edwards hypothesis.

One-dimensional spin systems at zero temperature
provide an adequate setting for the investigation of the dynamical role
of valleys, and especially for testing the Edwards hypothesis.
Indeed, the blocked configurations reached by zero-temperature dynamics
are truly metastable states with infinite lifetimes.
In an earlier work~\cite{us} we have addressed this question for Ising chains
evolving under irreversible zero-temperature dynamics,
where each spin may flip at most once before a blocked state is reached.
These dynamical models can be exactly mapped
onto random sequential adsorption problems~\cite{rsa},
for which analytical tools are available.

In the present work we pursue the investigation
of metastable states in Ising chains, by considering the richer
situation of zero-temperature Kawasaki (spin-exchange) dynamics~\cite{kawa}.
We allow the rate $W_0$ of diffusion processes
(constant-energy moves) to vary~\cite{cks}.
If the kinetic constraint $W_0=0$ is imposed~\cite{plk},
only irreversible processes are present~\cite{us}.
For the unconstrained dynamics $(W_0>0)$,
where diffusion processes are allowed, the dynamics is partly reversible.
Each spin only flips a finite number of times,
before the system globally reaches a blocked state after a finite time.
The statistics of this blocking time is studied in Section~\ref{s1}.
Our main goal is then again to test the Edwards hypothesis,
i.e., to evaluate various observables in the blocked configurations
reached by the dynamics,
starting from a random non-magnetized initial configuration.
We shall systematically compare the data of extensive numerical simulations,
presented in Section~\ref{s4},
with the predictions of the a priori approach, derived in Section~\ref{s3}.
Finally, some aspects of persistence,
in particular the distribution of the number of flips of a given spin,
are discussed in Section~\ref{s5}.

To be more specific,
we consider a ferromagnetic chain of Ising spins $\s_n=\pm1$,
whose Hamiltonian reads
\beq
\H=-\sum_n\s_n\s_{n+1}.
\label{hamf}
\eeq

In Kawasaki dynamics~\cite{kawa},
only pairs of opposite spins may be flipped $(+-\tot-+)$,
so that the magnetization is locally conserved.

For simplicity, we limit ourselves to Monte-Carlo dynamics
with random sequential updating, and we assume that the flipping rate
only depends on the energy difference $\delta\H$ involved in the proposed move.
Any given pair of opposite spins thus has a probability
$W_{\delta\H}$ of flipping per unit time, with
\be
\delta\H=2(\s_{n-1}\s_n+\s_{n+1}\s_{n+2})\in\{-4,0,4\}.
\ee
The requirement of detailed balance
with respect to the Hamiltonian~(\ref{hamf}) at temperature $T=1/\beta$
yields a single condition,
\be
\frac{W_4}{W_{-4}}=\e^{-4\beta},
\ee
upon the rates $W_4$, $W_0$, $W_{-4}$.

We furthermore restrict ourselves to zero-temperature dynamics,
hence $W_4=0$.
We choose time units such that $W_{-4}=1$, keeping $W_0$ as a free parameter.
The allowed moves and the corresponding rates are listed in Table~\ref{t1}.
The zero-temperature limits of the
Metropolis and heat-bath rules correspond respectively
to $W_0=1$ and $W_0=1/2$.
Hereafter we focus our attention on the range $0\le W_0\le1$.

\begin{table}[htb]
\begin{center}
\begin{tabular}{|c|c|c|c|c|}
\hline
$\delta\H$&type&name&rate&moves\\
\hline
$-4$&irreversible&condensation&1&$\matrix{-+-\,+\to--+\,+\cr+-+\,-\to++-\,-}$\\
\hline
0&reversible&diffusion&$W_0$&$\matrix{++-\,+\tot+-+\,+\cr-+-\,-\tot--+\,-}$\\
\hline
\end{tabular}
\caption{\small Allowed moves in zero-temperature Kawasaki dynamics.}
\label{t1}
\end{center}
\end{table}

\section{Statistics of blocking time}
\label{s1}

Let us first recall that for the kinetically constrained model $(W_0=0)$
a finite system consisting of~$N$ spins
reaches a blocked configuration after a finite blocking time $T_N\sim\ln N$,
which is the jamming time of the equivalent problem
of random sequential adsorption of hollow trimers~\cite{us}.
The blocked configurations are characterized by the property that
the spin patterns $+-+-\null$ and $-+-+\null$ are absent.
Equivalently, there are at most two consecutive unsatisfied bonds.

In the present case $(W_0>0)$
the system still gets trapped in a blocked configuration.
However the diffusive motion of free $+$ spins in domains of $-$ spins,
and vice-versa, is allowed.
Each free spin will eventually be annihilated,
by meeting either another free spin or one of the boundaries of the domain.
Blocked configurations of Kawasaki dynamics
are therefore characterized by the property that
the patterns $+-+\null$ and $-+-\null$ are absent.
Equivalently, isolated spins are absent, or unsatisfied bonds are isolated.

In order to understand the statistics of the blocking time $T_N$,
we consider first the regime $W_0\ll1$,
where the time scales of condensation and diffusion are well
separated~\cite{cks}.
The fast part of the dynamics, which occurs with unit rate,
is identical to the constrained dynamics considered in Refs.~\cite{us,plk}.
For intermediate times of order $1\ll t\ll1/W_0$,
the system is therefore approximately left in one of the final configurations
of the constrained dynamics.
The slow, diffusive part of the dynamics then takes place at rate $W_0\ll1$.
The late stages of the dynamics are governed by large domains,
on which a single free spin diffuses.
We assume that such large domains, of size $L\gg1$,
occur with an exponentially small probability
\be
f_\dif(L)\sim\exp(-L/\xi_\dif),
\ee
where $\xi_\dif$ is the relevant characteristic length.
The spin diffusion constant reads $D=W_0$ in our units.
For a random initial point,
the survival probability is known~\cite{trap} to decay as
$S(t;L)\approx(8L/\pi^2)\exp(-\pi^2W_0t/L^2)$.
The mean density of free spins at time~$t$ can therefore be estimated as
\beq
S(t)\approx\sum_L f_\dif(L)S(t;L)
\sim\int_0^\infty\exp\left(-\frac{L}{\xi_\dif}-\frac{\pi^2W_0t}{L^2}\right)
\label{sint}
L\,\d L.
\eeq
Evaluating the integral by the method of steepest descent,
we thus obtain stretched exponential decay for the density of free spins:
\beq
S(t)\sim t^{1/2}\,\exp\left(-\frac{(W_0t)^{1/3}}{A_\dif}\right),
\label{st}
\eeq
with
\beq
A_\dif=\left(\frac{4\xi_\dif^2}{27\pi^2}\right)^{1/3}.
\label{ad}
\eeq

This behavior, already emphasized in Ref.~\cite{cks},
is a general characteristic feature of diffusion processes
in the presence of random traps~\cite{trap}.
The exponent $1/3$ is related to the one-dimensional geometry;
it would read $d/(d+2)$ for trapping problems in higher spatial dimension $d$.
The above result should hold for the late stages $(W_0t\gg1)$
of Kawasaki dynamics, for any finite value of $W_0$.
The characteristic length $\xi_\dif$ is expected to depend smoothly on $W_0$.

For a finite system of~$N$ spins, the last free spin will be annihilated
at a time $T_N$ such that $NS(T_N)\sim1$,
hence $(W_0T_N)^{1/3}\sim A_\dif\ln N$.
More precisely, as the histories of spins diffusing
on different domains are statistically independent,
it can be argued along the lines of~\cite{us}
that $T_N$ is given according to extreme-value statistics~\cite{gumbel} as
\beq
(W_0T_N)^{1/3}\approx A_\dif\left(\,\ln\!\Bigl(N(\ln N)^{3/2}\Bigr)
+b+X_N\right).
\label{xdef}
\eeq
In this expression, the factor $(\ln N)^{3/2}$ takes account of the
$t^{1/2}$ prefactor in the survival probability~(\ref{st}),
while the effective constant~$b$ encompasses all subleading effects
that have been neglected,
and the random variable $X_N$ is distributed according to the Gumbel law
\beq
f(X)=\exp(-X-\e^{-X}).
\label{gum}
\eeq

\begin{figure}[htb]
\begin{center}
\includegraphics[angle=90,width=.7\linewidth]{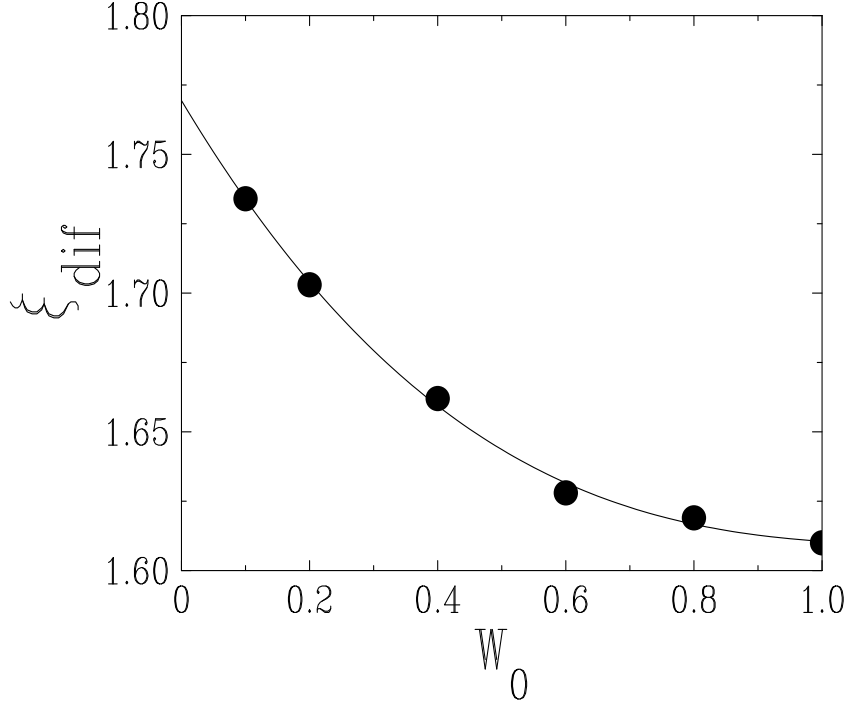}
\caption{\small
Plot of the characteristic length $\xi_\dif$,
extracted from the size dependence of $\mean{(W_0T_N)^{1/3}}$,
against the rate $W_0$.
Symbols: numerical data.
Errors are comparable to the symbol size.
Line: third-degree polynomial fit.}
\label{f1}
\end{center}
\end{figure}

These predictions have been checked against extensive numerical simulations,
performed according to the
zero-temperature Kawasaki dynamics summarized in Table~\ref{t1},
with random sequential updates and periodic boundary conditions,
starting from a random initial configuration with zero magnetization.
The dynamics is run until the system gets trapped in a blocked configuration.
The blocking time $T_N$ is recorded for each sample.
This measurement has been performed for many samples
($10^8$ spins in total for each value of $W_0$ and of $N$).
The mean $\mean{(W_0T_N)^{1/3}}$ is found to follow an almost perfect
linear law when plotted against $\ln(N(\ln N)^{3/2})$,
at least for~$N$ ranging from 50 to 3200,
whereas the same data plotted against $\ln N$ are bent in a significant way.
Using~(\ref{ad}), the slope of the latter plot yields
the value of the characteristic length $\xi_\dif$,
which is plotted in Figure~\ref{f1}, against $W_0$.
The error on this estimate, containing a systematic and a statistical part,
is roughly comparable to the symbol size.
The length $\xi_\dif$ exhibits a rather weak dependence on $W_0$,
decreasing from $\xi_\dif=1.77$ in the $W_0\to0$ limit
to $\xi_\dif=1.61$ for $W_0=1$.

\begin{figure}[htb]
\begin{center}
\includegraphics[angle=90,width=.7\linewidth]{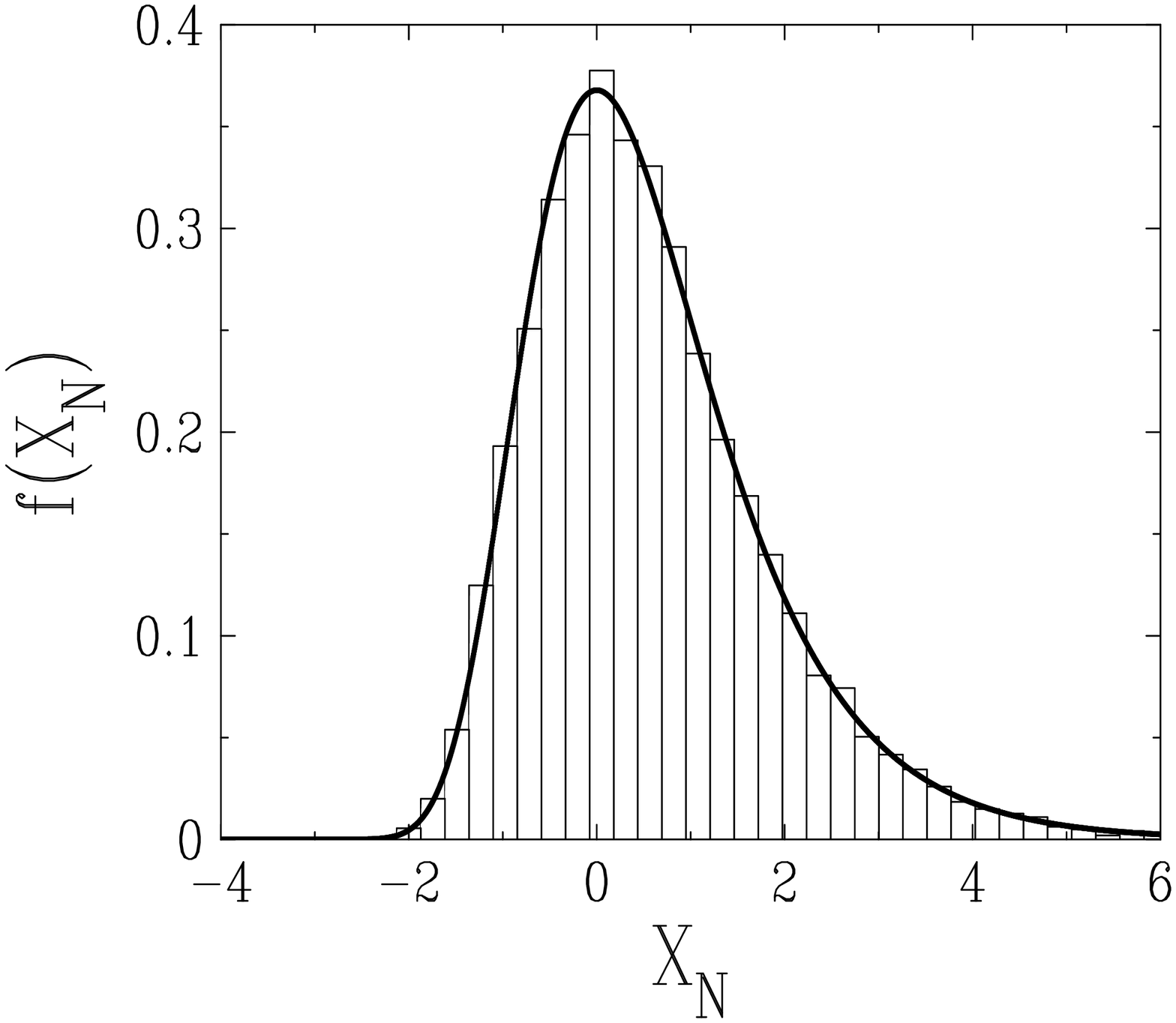}
\caption{\small
Distribution of the variable $X_N$ defined by~(\ref{xdef}).
Histogram: numerical data for $N=800$ (see text).
Thick line: limit Gumbel law~(\ref{gum}).}
\label{f2}
\end{center}
\end{figure}

Another confirmation of the above picture is provided by Figure~\ref{f2},
showing a histogram plot of the variable $X_N$
defined by~(\ref{xdef}), for 125,000 samples of size $N=800$, with $W_0=1$.
The parameter $A_\dif=0.339$, i.e., $\xi_\dif=1.61$,
is taken from the data of Figure~\ref{f1},
while the constant~$b$ is chosen by fitting the average $\mean{X_N}$.
A convincing agreement is found with the limit law~(\ref{gum}).

\section{Blocked states: a priori ensemble}
\label{s3}

As already stated, the blocked states of zero-temperature Kawasaki dynamics
are the configurations where isolated spins are absent.
Equivalently, unsatisfied bonds are isolated.

This section is devoted to the statistical description of the blocked
configurations taken with equal a priori weights (Edwards approach).
We shall distinguish the {\it full} ensemble
of all the blocked configurations, irrespective of their energy,
the {\it restricted} ensemble of blocked configurations
with prescribed energy density $E$,
and the {\it canonical} ensemble of blocked configurations,
obtained by fixing the parameter $\beta$ conjugate to the energy density $E$
(see below).

\subsection{Statistics of energy and configurational entropy}

For a finite chain of~$N$ spins,
we first investigate the number $\N(N;E)$ of blocked configurations
with prescribed energy density $E$.
This number can be evaluated
by an elementary combinatorial reasoning~\cite{us}.
For the sake of generality,
we prefer to resort to the transfer-matrix formalism.
We introduce the partition function $Z_N(\beta)$,
defined as a sum over all the blocked configurations $\C=\{\s_n\}$
of the Boltzmann weight associated with the Hamiltonian~(\ref{hamf}).
We have
\beq
Z_N(\beta)=\sum_\C\e^{-\beta\H(\C)}\approx\int\N(N;E)\,\e^{-\beta NE}\,\d E.
\label{zn}
\eeq
In this framework, the parameter $\beta$ can be positive or negative,
and it is not related to physical temperature.
The transfer matrix is then a very useful tool.
Indeed the partition functions $Z_N^\alpha$,
labeled by the prescribed value $\alpha=(\s_{N-1},\s_N)$
of the last two spins, obey the recursion relation
\be
\pmatrix{Z_{N+1}^{++}\cr Z_{N+1}^{+-}\cr Z_{N+1}^{-+}\cr Z_{N+1}^{--}}
=\T\pmatrix{Z_N^{++}\cr Z_N^{+-}\cr Z_N^{-+}\cr Z_N^{--}},
\ee
where the $4\times4$ transfer matrix $\T$ reads
\be
\T=\pmatrix{\e^\beta&0&\e^\beta&0\cr\e^{-\beta}&0&0&0\cr
0&0&0&\e^{-\beta}\cr 0&\e^\beta&0&\e^\beta}.
\ee

The characteristic polynomial of $\T$ factors as
$\det(\lam\1-\T)=(\lam^2-\e^{\beta}\lam-1)(\lam^2-\e^{\beta}\lam+1)$.
The first (resp.~the second) factor yields eigenvalues $\lam_1$, $\lam_2$
(resp.~$\lam_3$, $\lam_4$), with
\be
\lam_{1,2}=\frad{1}{2}\left(\e^{\beta}\pm\sqrt{\e^{2\beta}+4}\right),\qquad
\lam_{3,4}=\frad{1}{2}\left(\e^{\beta}\pm\sqrt{\e^{2\beta}-4}\right).
\ee
The left and right eigenvectors $\langle L_a\vert$ and $\vert R_a\rangle$
are even (resp.~odd) for $a=1,2$ (resp.~$a=3,4$)
under the spin symmetry $+\tot-$.
We assume that they are normalized so that
$\langle L_a\vert R_b\rangle=\delta_{ab}$.
Their explicit expressions will not be needed hereafter.

For large $N$, we have $Z_N(\beta)\sim\lam_1^N$,
as $\lam_1$ is the largest eigenvalue.
Using~(\ref{zn}), we obtain an exponential law of the form~(\ref{ne})
for $\N(N;E)$, where the a priori configurational entropy $S\ap(E)$
is related to $\ln\lam_1(\beta)$ by a Legendre transform:
\be
S\ap(E)-\ln\lam_1(\beta)=\beta E,
\qquad E=-\frac{\d\ln\lam_1}{\d\beta},
\qquad\beta=\frac{\d S\ap}{\d E}.
\ee
Explicitly, we have
\beq
E=-\frac{\e^\beta}{\sqrt{\e^{2\beta}+4}},\qquad
\e^\beta=\frac{-2E}{\sqrt{1-E^2}},
\label{eb}
\eeq
and the a priori entropy reads
\be
S\ap(E)=E\ln(-2E)+\frac{1-E}{2}\ln(1-E)-\frac{1+E}{2}\ln(1+E).
\ee
This entropy is non-zero for $-1<E<0$.
It takes its maximal value
\be
S\st=\ln\Phi=0.481212,
\ee
where $\Phi=(1+\sqrt{5})/2$ is the golden mean,
for
\beq
E\st=-\frac{1}{\sqrt{5}}=-0.447214,
\label{est}
\eeq
corresponding to $\beta=0$.
Equation~(\ref{est}) therefore yields the typical a priori energy density
of a blocked configuration.

The result~(\ref{ne}) can be recast as follows.
Consider the full ensemble of all the blocked configurations,
irrespective of their energy.
The probability of observing, in that ensemble,
a blocked configuration with energy density~$E$ reads
\beq
P\ap(E)\sim\exp(-N\S\ap(E)),\qquad\S\ap(E)=S\st-S\ap(E).
\label{sap}
\eeq
The function $\S\ap(E)$ vanishes quadratically as
\be
\S\ap(E)\approx c\left(E-E\st\right)^2,\qquad c=\frac{5\sqrt{5}}{8}.
\ee
The bulk of the a priori distribution of $E$
is therefore a narrow Gaussian around $E\st$,
whose rescaled variance asymptotically reads
\be
N\var E=N(\mean{E^2}-E\st^2)\toinf\frac{1}{2c}=\frac{4\sqrt{5}}{25}=0.357771.
\ee

The above results also allow to determine the higher cumulants
of the energy in the canonical ensemble (at fixed parameter $\beta$).
One has indeed
\be
\frac{1}{N}\ln\mean{\e^{sNE}}_\beta
=\frac{1}{N}\ln\frac{Z_N(\beta-s)}{Z_N(\beta)}
\toinf\ln\frac{\lam_1(\beta-s)}{\lam_1(\beta)}.
\ee
By expanding this result as a power series in $s$,
we obtain explicit expressions for the cumulants $\cu{E^k}$
as a function of $\beta$.

As far as mean quantities are concerned,
the microcanonical or restricted ensemble (fixed energy density $E$)
and the canonical one (fixed conjugate parameter~$\beta$) are equivalent.
It is therefore justified to recast canonical results
in terms of the prescribed value~$E$ of the mean energy, using~(\ref{eb}).
Generalizing this equivalence prescription to the cumulants of the energy,
we obtain
\beqa
N\var{E}=N\cu{E^2}&\toinf&-E(1-E^2),\nonumber\\
N^2\cu{E^3}&\toinf&-E(1-E^2)(3E^2-1),\label{ecu}\\
N^3\cu{E^4}&\toinf&-E(1-E^2)(15E^4-12E^2+1).\nonumber
\eeqa
%The energy cumulants are definitely not mean quantities.
The results~(\ref{ecu}) can be given the following interpretation.
Fixing $\beta$ amounts to fixing the extensive part $NE$ of the energy,
while the equivalence prescription
is a natural Ansatz to describe the fluctuations of its non-extensive part.

\subsection{Spin correlation function}

The spin correlation function $C_n=\mean{\s_0\s_n}$ in the canonical
a priori ensemble can also be evaluated by the transfer-matrix method.
In the bulk of an infinitely long chain, and for $n\ge0$, we have
\beq
C_n=\frad{\langle L_1\vert\Sp\T^n\Sp\vert R_1\rangle}{\lam_1^n}
=\sum_a\langle L_1\vert\Sp\vert R_a\rangle\langle L_a\vert\Sp\vert R_1\rangle
\left(\frac{\lam_a}{\lam_1}\right)^n,
\label{ctm}
\eeq
where
\be
\Sp=\pmatrix{1&0&0&0\cr 0&-1&0&0\cr 0&0&1&0\cr 0&0&0&-1}
\ee
is the spin operator.

Because of symmetry,
only the eigenvectors with $a=3,4$ contribute to the above sum.
The values $C_0=1$ and $C_1=-E$ allow
to determine the products of matrix elements entering~(\ref{ctm}),
without knowing the eigenvectors explicitly.
The following alternative reasoning can also be used.
Being a linear combination of $(\lam_3/\lam_1)^n$ and $(\lam_4/\lam_1)^n$,
$C_n$ can be shown to obey the three-term recursion relation
\be
(1-E)C_{n+2}+2EC_{n+1}+(1+E)C_n=0.
\ee
The initial values $C_0=1$ and $C_1=-E$ are therefore again sufficient
to determine the correlation function for all values of the distance $n$.

For $-1/\sqrt{2}\le E\le0$,
which contains the range of final energies reached by the dynamics,
$\lam_{3,4}=\exp(\pm\i Q)$ are complex numbers with unit modulus, with
\beq
\tan Q=\frac{\sqrt{1-2E^2}}{-E}\qquad(0\le Q\le\pi/2).
\label{qap}
\eeq
We are thus led to the expression
\beq
C_n=\left(\frac{1+E}{1-E}\right)^{n/2}
\left(\cos nQ+\frac{E^2}{\sqrt{1-2E^2}}\,\sin nQ\right),
\label{cap}
\eeq
for $n\ge0$.
In the a priori ensemble,
the spin correlation function therefore exhibits an exponential decay,
of the form $\exp(-n/\xi_\spin)$, modulated by oscillations at wavevector~$Q$.
Both the correlation length
\beq
\xi_\spin=\frad{2}{\ln\frad{1-E}{1+E}}
\label{xiap}
\eeq
and the wavevector $Q$, given by~(\ref{qap}), depend continuously on energy.

We finally quote for further reference the value of the reduced susceptibility
\beq
\chat=\sum_{n=-\infty}^\infty C_n.
\label{chat}
\eeq
After some algebra, we obtain the simple result
\beq
\chat=\frac{-E(1-E)}{1+E}.
\label{chatap}
\eeq

\subsection{Distribution of domain sizes}

Another characteristic feature of blocked states
is the distribution of domain sizes $f(\ell)$,
defined as the probability that a given domain
consists exactly of $\ell$ consecutive parallel spins.
Since isolated spins are absent, domains have at least size two ($\ell\ge2$).

In the a priori ensemble, the distribution $f(\ell)$
can again be evaluated by the transfer-matrix method.
Indeed $\rho(\ell)$, the density (per unit length) of domains consisting of
exactly $\ell$ spins,
admits an expression similar to the middle side of~(\ref{ctm}), namely
\be
\rho(\ell)
=\frad{\langle L_1\vert\P\,\e^{(\ell-2)\beta}\vert R_1\rangle}{\lam_1^\ell}
=\langle L_1\vert\P\vert R_1\rangle
\;\frac{1+E}{1-E}\left(\frac{-2E}{1-E}\right)^{\ell-2},
\ee
where
\be
\P=\,(\mid\!+-\rangle\langle-+\!\mid)+(\mid\!-+\rangle\langle+-\!\mid)\,
=\pmatrix{0&0&0&0\cr 0&0&1&0\cr 0&1&0&0\cr 0&0&0&0}
\ee
is the appropriate domain boundary operator.
The explicit expression of the matrix element is not needed,
as it can be fixed by normalization.
The probability distribution $f(\ell)$ of domain sizes indeed
reads $f(\ell)=\rho(\ell)/\rho$, where
$\rho$ is the total density of domains (or equivalently, of domain walls).

We thus obtain the geometric (i.e., discrete exponential) probability
distribution
\beq
f(\ell)=\frac{1+E}{1-E}\left(\frac{-2E}{1-E}\right)^{\ell-2}
\label{fellap}
\eeq
for $\ell\ge2$, and consistently
\beq
\rho=\sum_{\ell=2}^\infty\rho(\ell)=\frac{1}{\mean{\ell}}=\frac{1+E}{2}
=\langle L_1\vert\P\vert R_1\rangle.
\label{dens}
\eeq
The characteristic length of~(\ref{fellap}),
\beq
\xi_\dom=\frad{1}{\ln\frad{1-E}{-2E}},
\label{xiclap}
\eeq
is in general different from the spin correlation length $\xi_\spin$
of~(\ref{xiap}), except for the typical value of energy~(\ref{est}),
where $\xi_\spin=\xi_\dom=1/S\st$.

The mean of the domain size distribution~(\ref{fellap})
agrees with~(\ref{dens}), while its variance reads
\be
\var\ell=\sum_{\ell=2}^\infty\ell^2 f(\ell)-\mean{\ell}^2
=\frac{-2E(1-E)}{(1+E)^2}.
\ee

Consider now a large sample of~$N$ spins,
in the canonical a priori ensemble (fixed parameter $\beta$).
The number~$M$ of domains in the sample is such that $N=\ell_1+\cdots+\ell_M$
is the sum of~$M$ independent variables distributed according to $f(\ell)$,
neglecting boundary effects.
The expected number of domains and its variance are given by
\beq
\frac{\mean{M}}{N}\toinf\frac{1}{\mean{\ell}}=\frac{1+E}{2},\qquad
\frac{\var{M}}{N}\toinf\frac{\var{\ell}}{\mean{\ell}^3}=\frac{-E(1-E^2)}{4}.
\label{varm}
\eeq
One has therefore
\beq
N\var{E}=\frac{4\var{M}}{N}\toinf\frac{4\var{\ell}}{\mean{\ell}^3},
\label{evarap}
\eeq
so that the second result of~(\ref{varm})
agrees with expression~(\ref{ecu}) for $N\var{E}$.

\section{Blocked states: dynamics}
\label{s4}

In this section we compare the predictions of the a priori approach,
derived in Section~\ref{s3}, to the results of numerical simulations
concerning the blocked configurations reached by the dynamics.
We have used the rules of the zero-temperature Kawasaki dynamics summarized in
Table~\ref{t1}, with random sequential updates,
starting from a random non-magnetized initial configuration.

\subsection{Mean energy}

The first and the simplest quantity to be measured is the mean energy $E$
of the blocked configurations reached by the dynamics.

Let us start with a reminder of the kinetically constrained model $(W_0=0)$.
For a random initial configuration, the two kinds of allowed defects, namely
isolated unsatisfied bonds (domain walls)
and domains of two unsatisfied bond (isolated spins)
occur respectively with the following densities in blocked
configurations~\cite{us,plk}:
\be
q_1=\frac{1}{2}\left(1-\e^{-5/4}
-\e^{-9/4}\int^{3/2}_1\e^{y^2}\,\d y\right)=0.219704,\qquad
q_2=\frac{\e^{-5/4}}{4}=0.071626,
\ee
so that the mean energy of blocked configurations reads
\beq
E_0=-1+2q_1+4q_2=-\e^{-9/4}\int^{3/2}_1\e^{y^2}\,\d y=-0.274087.
\label{e0}
\eeq

In the present case $(W_0\ne0)$, the mean energy~$E$ of
blocked states is expected to be below this number.
Indeed the diffusive moves can only help relaxing more efficiently
the energy excess of the disordered initial state.

\begin{figure}[htb]
\begin{center}
\includegraphics[angle=90,width=.7\linewidth]{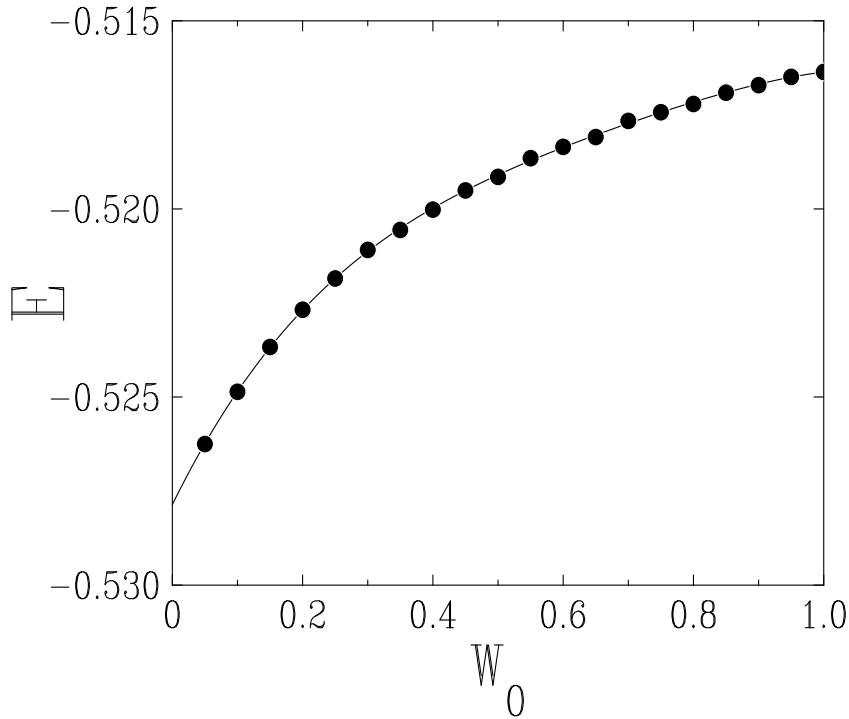}
\caption{\small
Plot of the mean final energy~$E$ against the rate $W_0$.
Symbols: numerical data.
Statistical errors are smaller than the symbol size.
Line: fifth-degree polynomial fit, yielding the extrapolated value~(\ref{e00})
in the $W_0\to0$ limit.}
\label{f3}
\end{center}
\end{figure}

Figure~\ref{f3} shows a plot of the mean final energy $E$
against the diffusive rate $W_0$.
Each data point corresponds to $10^8$ spins in total.
We have checked that no appreciable size dependence is to be observed.
The final energy is found to be well below~(\ref{e0}),
and below the typical a priori value~(\ref{est}).
It only exhibits a very weak dependence on the rate $W_0$,
increasing from the extrapolated minimum value
\beq
E(0)=-0.5279
\label{e00}
\eeq
in the $W_0\to0$ limit to the maximum value
\beq
E(1)=-0.51633
\label{e}
\eeq
for $W_0=1$.

The regime $W_0\ll1$ again deserves some more attention.
The mean energy has a discontinuity at $W_0=0$.
Indeed the $W_0\to0$ limit of the energy, $E(0)$ given by~(\ref{e00}),
is different from that of the constrained dynamics, $E_0$ given by~(\ref{e0}).
This discontinuity can be analyzed as follows.
In the diffusive part of the dynamics,
free spins annihilate by meeting either each other or domain walls.
The collision of a single free spin with a domain wall
relaxes 4 units of energy,
while the coalescence between $n\ge2$ spins within a domain
relaxes $4(n-1)$ units of energy, i.e., $4(n-1)/n$ per spin.
If all the free spins were annihilated in meeting domain walls,
the final energy would assume the value $E=E_0-4p_2$.
The efficiency of the diffusive relaxation mechanism
can therefore be characterized by the ratio
\be
\eta=\frac{E_0-E(0)}{4p_2}=1-\sum_{n\ge2}\frac{\Pi_n}{n},
\ee
where $\Pi_n$ is the probability that a given free spin
gets annihilated in a coalescence of $n\ge2$ spins.
The extrapolated value~(\ref{e00}) of the final energy
yields a rather high efficiency: $\eta=0.886$.

In view of the weak dependence of the final energy
on the rate $W_0$ (see Figure~\ref{f3}),
hereafter we restrict the numerical analysis to the case $W_0=1$
of zero-temperature Metropolis dynamics.

\subsection{Higher cumulants of energy}

We now turn to the statistics of the energy of the blocked configurations,
besides its mean value studied above.
In analogy with the a priori estimate~(\ref{sap}),
the final energy is expected to obey a large-deviation formula of the type
\be
P(E)\sim\exp(-N\S(E)),
\ee
with $\S(E)$ being the dynamical entropy.
This formula implies that the cumulants of the total energy scale
as $\cu{(NE)^k}\sim N$, just as in usual equilibrium situations.

Instead of measuring the whole function $\S(E)$,
which would require a quite extensive numerical effort,
we have measured the first four cumulants of the energy
of the blocked configurations.
Table~\ref{t2} gives the measured values of the scaled
energy cumulants for $W_0=1$ and a random initial configuration.
For comparison we also list the predictions of the full
a priori ensemble~(\ref{ecu}) for $E=E\st$ of~(\ref{est}),
and of the restricted a priori ensemble,
obtained by inserting into~(\ref{ecu})
the observed value~(\ref{e}) of the mean energy.
The simulations have been performed on samples of various sizes
ranging from $N=50$ to 200, having $10^{10}$ spins in total.
No systematic size dependence is observed.
Statistical errors can be estimated to be of the order
of one unit of the least significant digit.
Both a priori schemes perform very unequally in predicting
the energy cumulants.
The variance is rather accurately predicted by both schemes,
which perform equally poorly for the third cumulant,
while the restricted scheme is definitely better for the fourth cumulant.

\begin{table}[htb]
\begin{center}
\begin{tabular}{|c|c|c|c|}
\hline
scaled cumulant&numerical result&full a priori&restricted a priori\\
\hline
$\cu{E}=E$&$-0.51633$&$-0.44721$&$-0.51633$\\
$N\cu{E^2}=N\var{E}$&{\hskip 1.2mm}$0.3446$&{\hskip 3.3mm}$0.35777$&{\hskip
3.3mm}$0.37868$\\
$N^2\cu{E^3}$&{\hskip -0.9mm}$0.031$&$-0.14311$&$-0.07582$\\
$N^3\cu{E^4}$&{\hskip -6.3mm}$-0.46$&$-0.28622$&$-0.42906$\\
\hline
\end{tabular}
\caption{\small First four scaled cumulants of the energy
of the blocked configurations.
Comparison between numerical results
and predictions of the full and restricted a priori ensembles (see text).}
\label{t2}
\end{center}
\end{table}

\subsection{Spin correlation function}

We now turn to the spin correlation function
$C_n=\mean{\s_0\s_n}$ in the blocked configurations.
This quantity has two remarkable properties.
First, since isolated spins are not allowed in blocked configurations,
we have $C_1=1-2\rho=-E$ and $C_2=1-4\rho$,
where $\rho$ is the density of domain walls~(\ref{dens}), hence
\beq
C_2=2C_1-1.
\label{c1c2}
\eeq
Second, the total magnetization $M=\sum_n\s_n$ is exactly conserved
in Kawasaki dynamics.
Furthermore, for a homogeneous non-magnetized state,
we have $\mean{M^2}=N\chat$, with the definition~(\ref{chat}).
The quantity $\chat$ therefore has the same value in the initial
and final configurations, i.e., for a random initial state,
\beq
\chat=\sum_{n=-\infty}^\infty C_n=1.
\label{chat1}
\eeq

\begin{figure}[htb]
\begin{center}
\includegraphics[angle=90,width=.7\linewidth]{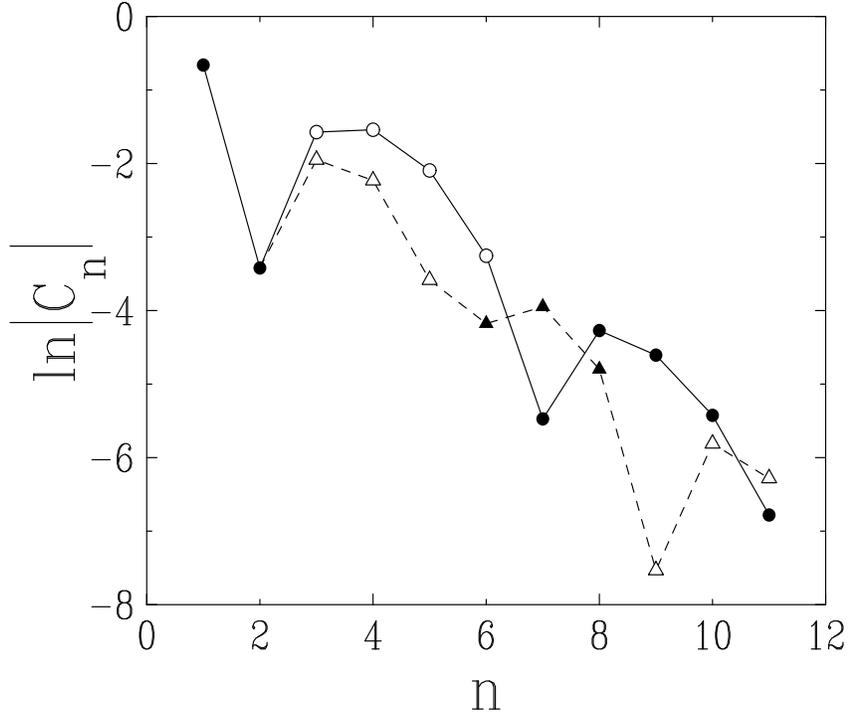}
\caption{\small
Logarithmic plot of the absolute value of the
spin correlation function $\abs{C_n}$, against distance $n$.
Full (open) symbols show positive (negative) correlations.
Circles and full line: numerical data.
Statistical errors are smaller than the symbol size.
Triangles and dashed line: prediction~(\ref{cap}) of restricted
a priori ensemble at energy~(\ref{e}).}
\label{f4}
\end{center}
\end{figure}

Figure~\ref{f4}
shows a comparison between numerical data (circles and full line),
corresponding to $3\cdot10^9$ spins in total,
and the prediction~(\ref{cap}) of the restricted a priori ensemble
at the observed mean energy~(\ref{e}) (triangles and dashed line).
The a priori prediction and numerical data coincide
both for $n=1$ (by construction, as~$E$ has been imposed)
and for $n=2$ (as a consequence of~(\ref{c1c2})).
For larger distances, the spin correlations
oscillate in sign and fall off exponentially,
in qualitative agreement with the a priori prediction.
Both the period of oscillations and the decay length
are observed to be slightly larger than those of the a priori ensemble.
These observations remain however at a qualitative level.

We also notice that the a priori ensemble
fails to reproduce the identity~(\ref{chat1}).
Indeed~$\chat$ of~(\ref{chatap}) equals unity for $E=1-\sqrt{2}=-0.414214$,
a value which neither agrees with the most probable
a priori energy $E\st$ of~(\ref{est}),
nor with the observed mean energy~(\ref{e}).
Conversely, the prediction for the correlation function
of the restricted a priori ensemble at energy~(\ref{e}),
plotted in Figure~\ref{f4},
has $\chat=1.6192$, which is significantly different from unity.

\subsection{Distribution of domain sizes}

\begin{figure}[htb]
\begin{center}
\includegraphics[angle=90,width=.7\linewidth]{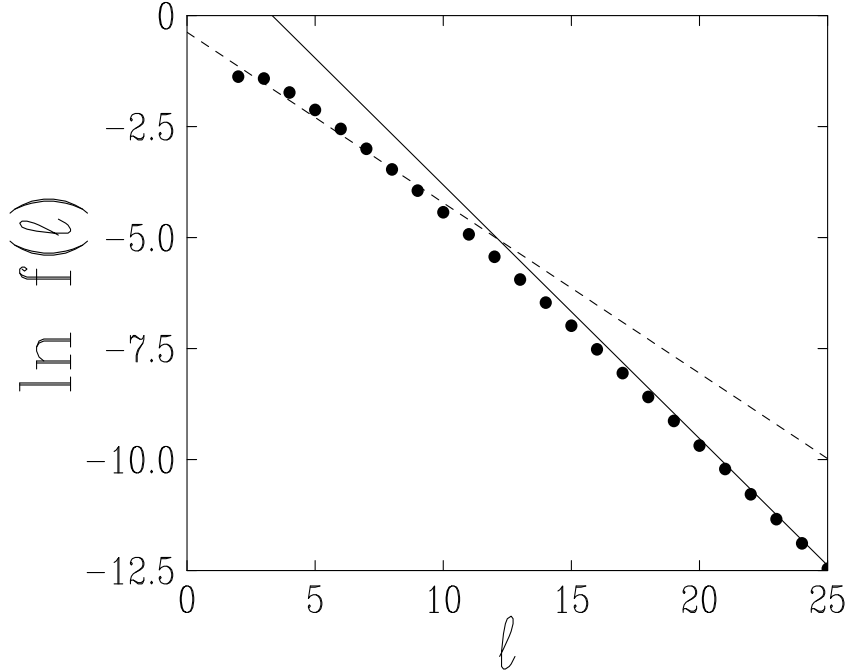}
\caption{\small
Logarithmic plot of the probability distribution $f(\ell)$
against domain size $\ell$.
Circles: numerical data.
Statistical errors are much smaller than the symbol size.
Dashed straight line: prediction~(\ref{fellap}) of a priori ensemble
at energy~(\ref{e}).
Full straight line: guide to the eye with inverse slope $\xi_\dom=1.75$.}
\label{f5}
\end{center}
\end{figure}

We have determined the distribution of domain sizes $f(\ell)$
by means of extensive numerical simulations, with a random initial state.
Figure~\ref{f5} shows our data corresponding to $3\cdot10^9$ spins in total.
The observed distribution is not exponential,
at variance with the prediction~(\ref{fellap}) of the a priori ensemble.
The full line suggests an exponential asymptotic fall-off of the
distribution, with a characteristic length $\xi_\dom=1.75$.
This length is close to the characteristic length
of domains with a single diffusive spin, $\xi_\dif=1.61$,
plotted in Figure~\ref{f1}.
As a matter of fact,
the equality $\xi_\dom=\xi_\dif$ is expected to hold in the $W_0\to0$ limit.
The prediction~(\ref{fellap}) of the restricted a priori ensemble
at the observed mean energy~(\ref{e})
is shown as a straight dashed line, whose inverse slope
is the prediction $\xi_\dom=2.603$~(\ref{xiclap}) of the a priori approach.

Another noticeable difference with the prediction of the a priori approach
is that the sizes of successive domains in the final states of the dynamics
are not independent.
Indeed, if they were so, the relation~(\ref{evarap}) would hold,
while the data yield $4\var{\ell}/\mean{\ell}^3=0.2790$,
a number significantly below the observed variance $N\var{E}=0.3446$,
listed in Table~\ref{t2}.
Hence domain sizes exhibit a weak but definite {\it positive} correlation.
Domain sizes which are neither statistically independent
nor exponentially distributed have also been observed recently~\cite{bfs}
in spin chains undergoing tapping dynamics.

\section{Features of persistence}
\label{s5}

We now turn to features related to the whole history of a single given spin,
say $\s_0$, the spin situated at the origin.
This kind of problems belongs to the realm of persistence.
In the present context,
the central quantity is $\nu$, the total number of times $\s_0$ flips
during the history of the sample, before a blocked configuration is reached.
The number $\nu$ of spin flips is finite with unit probability
in the limit of a large system.
Zero-temperature Kawasaki dynamics is therefore of type~${\cal F}$
in the classification of~\cite{ns}.
Furthermore $\nu$ is random, as it depends both on the initial configuration
and on the history of the system.
We are therefore interested in the distribution $p_\nu$ of the number of flips,
which is expected to have a well-behaved limit when
$\s_0$ is deep inside a large enough sample.

Before presenting numerical data,
we first predict the main salient features of the distribution $p_\nu$,
following the lines of Section~\ref{s1}.
Consider again a large domain of $L\gg1$ spins,
containing the origin, on which a single free spin diffuses.
The spin $\s_0$ flips twice each time the free spin traverses the origin.
We are thus led to the following effective problem.

Consider a random walker in the interval $-L_1<n<L_2$,
with absorbing boundaries at $n=-L_1$ and $n=L_2$.
The walker starts from the origin ($n=0$).
The probability that the walker returns to the origin before being
absorbed by either boundary reads
\beq
P_\ret(L_1,L_2)=1-\frac{1}{2}\left(\frac{1}{L_1}+\frac{1}{L_2}\right).
\label{pgamb}
\eeq
This formula relies on a well-known result in
the gambler's ruin problem~\cite{feller}.
Indeed, suppose that the walker's first jump is to the right
(resp.~to the left),
and consider $0\le n\le L_2$ (resp.~$0\le -n\le L_1$) as the gambler's wealth.
Then the ruin probability reads $P=1-1/L_2$ (resp.~$P=1-1/L_1$).
The expression~(\ref{pgamb}) is the arithmetical mean of both ruin
probabilities.

The probability that $\s_0$ flips an even number $\nu=2k\gg1$ of times
therefore approximately reads
$p_{2k}(L_1,L_2)\sim(P_\ret(L_1,L_2))^k(1-P_\ret(L_1,L_2))$.
Hence the distribution of the number of flips can
be estimated, in analogy with~(\ref{sint}), as
\bea
p_{2k}&\approx&\sum_{L_1,L_2}(L_1+L_2-1)\,f_\dif(L_1+L_2-1)
\,P_\ret(L_1,L_2)^k(1-P_\ret(L_1,L_2))\\
&\sim&\left(\int_0^\infty\exp\left(-\frac{L}{\xi_\dif}-\frac{k}{2L}\right)
\,\d L\right)^2.
\eea
The integral entering this expression closely resembles
that entering~(\ref{sint}).
For $k\gg1$ it is legitimate to use the steepest-descent method.
The saddle point lies at $L_c=\sqrt{k\xi_\dif/2}$.
We thus obtain the stretched exponential law
\beq
p_{2k}\sim\sqrt{k}\,\exp\!\left(-2\sqrt{\frac{2k}{\xi_\dif}}\right)
\label{gp}
\eeq
for the distribution of the number of spin flips,
provided $\nu=2k$ is a large even number.

The occurrence of odd numbers $\nu=2k+1$ of spin flips can also
be explained in the above framework,
if the spin $\s_0$ is either situated at an endpoint of a domain,
or involved in a coalescence event between two free spins.
Both effects are expected to scale as the inverse of the domain size $L$.
This leads us to predict that odd values of $\nu$ are suppressed
by a factor of order $1/L_c$, i.e.,
\beq
\frac{p_{2k+1}}{p_{2k}}\approx\frac{a}{\sqrt{k\xi_\dif}}.
\label{gi}
\eeq

\begin{figure}[htb]
\begin{center}
\includegraphics[angle=90,width=.7\linewidth]{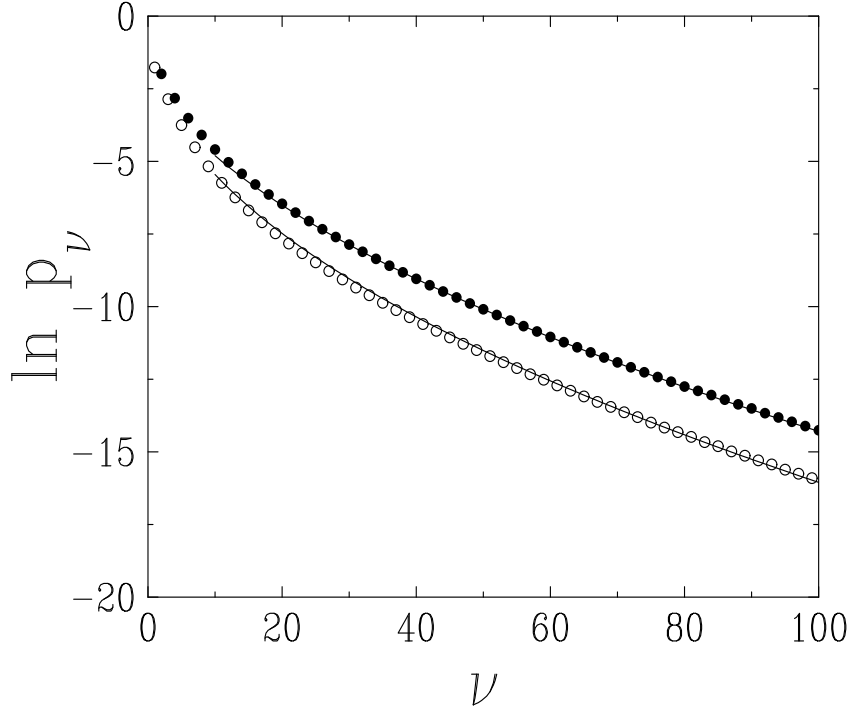}
\caption{\small
Logarithmic plot of the distribution $p_\nu$ of number of flips.
Full symbols: data for $\nu=2k$ even.
Open symbols: data for $\nu=2k+1$ odd.
Statistical errors are much smaller than the symbol size.
Full lines: common fit described in the text.}
\label{f6}
\end{center}
\end{figure}

Figure~\ref{f6} shows a logarithmic plot of
the distribution $p_\nu$ of the number of flips.
The simulations again concern samples of various sizes
having $2\cdot 10^{10}$ spins in total, with $W_0=1$.
Even numbers of spin flips $\nu=2k$ (full symbols)
are clearly more frequent than odd numbers $\nu=2k+1$ (open symbols),
especially for large values of $k$.
From a quantitative viewpoint, the full lines on Figure~\ref{f6}
show a common fit of the numerical data for $\nu>10$
according to the asymptotic predictions~(\ref{gp}),~(\ref{gi}).
We thus obtain $2/\sqrt{\xi_\dif}\approx1.55$,
in agreement with the data of Figure~\ref{f1},
$\xi_\dif\approx1.61$, i.e., $2/\sqrt{\xi_\dif}\approx1.57$.
We also obtain $a\approx1.5$, albeit with a large uncertainty.

Besides the above analysis of the regime of large numbers of spin flips,
due to long surviving free spins,
our data yield yet other interesting informations.
First, the persistence probability,
i.e., the probability for a spin to never flip,
has the following value:
\be
p_0=0.44739.
\ee
Then, the probability that a spin flips an even or an odd number of times reads
\be
P_\ev=\sum_{k\ge0}p_{2k}=0.72302,\qquad
P_\od=\sum_{k\ge0}p_{2k+1}=1-P_\ev=0.27698.
\ee
These figures can be related to the overlap between
a random initial configuration and the corresponding final one.
We have indeed
\be
Q=\mean{\s_0(0)\s_0(\infty)}=\sum_{\nu\ge0}(-1)^\nu p_\nu=P_\ev-P_\od=0.44604.
\ee

The mean number of spin flips reads
\be
\mean{\nu}=\sum_{\nu\ge0}\nu\,p_\nu=2.06916.
\ee
This quantity can be used to determine the fractions
$f_\cond$ of condensation moves
and $f_\dif$ of diffusive moves in a typical history,
starting from a random initial configuration.
Indeed, on the one hand, a condensation move lowers the total energy
by four units, while a diffusive moves leaves it unchanged
(see Table~\ref{t1}).
On the other hand, any move involves exactly two spin flips.
We have therefore, using~(\ref{e}),
\be
f_\cond=1-f_\dif=-\frac{E}{2\mean{\nu}}=0.12477.
\ee

To close up, let us compare the above results
to the case of constrained Kawasaki dynamics ($W_0=0$)~\cite{us,plk},
where only condensation moves are allowed.
In this situation, the above quantities
can be simply related to the final energy $E_0$ of~(\ref{e0}).
Indeed every spin flips at most once,
so that only $p_0$ and $p_1=1-p_0$ are non-zero, and $f_\cond=1$.
Starting again from a random configuration, we are left with
\be
2(1-p_0)=1-Q=2\mean{\nu}=-E_0=0.274087.
\ee

\section{Discussion}

First, we wish to emphasize the richness of the zero-temperature
dynamics of the ferromagnetic Ising chain.
There are indeed four different natural kinds of dynamics,
summarized in Table~\ref{t3}.
Only Glauber dynamics gives rise to a bona fide coarsening dynamics,
obeying dynamical scaling with a typical domain size
growing as $L(t)\sim\sqrt{t}$.
With the three other dynamics,
the system is left in a metastable configuration after a
relatively short blocking time.
The constrained Glauber and Kawasaki dynamics
have been analyzed in our previous work~\cite{us}.
Each spin may flip at most once, before a global blocked state is reached.
The blocking time scales with the number of spins as $\ln N$.
These models can be exactly mapped onto
the random sequential addition problem of dimers and hollow trimers,
respectively, hence allowing an analytical treatment.

\begin{table}[htb]
\begin{center}
\begin{tabular}{|l|l|l|l|}
\hline
$\matrix{\hbox{conserved}\hfill\cr\hbox{magnetization}}$
&$\matrix{\hbox{diffusive}\cr\hbox{moves}\hfill}$
&dynamics&behavior\\
\hline
no&yes&Glauber~\cite{glau}&coarsening~\cite{bray}\\
no&no&constrained Glauber~\cite{cg}&metastability~\cite{us}\\
yes&yes&Kawasaki~\cite{kawa}&metastability~[this work]\\
yes&no&constrained Kawasaki~\cite{plk}&metastability~\cite{us}\\
\hline
\end{tabular}
\caption{\small Four different zero-temperature dynamics of
the ferromagnetic Ising chain, with appropriate references (brackets).}
\label{t3}
\end{center}
\end{table}

The last case not considered so far from the viewpoint of metastability,
Kawasaki dynamics, has been the subject of this work.
Because of the diffusive moves, the dynamics is only partly irreversible.
This novel feature makes the Kawasaki problem both richer
and hopefully closer to more realistic situations.
The number of flips of a given spin, although finite with probability one,
may be arbitrarily large.
The blocking time grows as $(\ln N)^3$.
On the other hand, analytical tools being no longer available,
we needed to have recourse to numerical simulations.

The present work
demonstrates that there is no good a priori statistical description
of the metastable states reached by Kawasaki dynamics.
In other words, the Edwards hypothesis is invalidated,
at least as an exact prescription.
Systematic differences are indeed observed
between numerical results and a priori predictions,
especially in the pattern of spin correlations~(Figure~\ref{f4})
and in the distribution of domain sizes~(Figure~\ref{f5}).
The latter are neither statistically independent nor exponentially distributed.

We also want to underline that the effective trapping description
of the late stages of the zero-temperature Kawasaki dynamics,
already emphasized in Ref.~\cite{cks},
appears to yield quantitative predictions
for several novel physical quantities,
including the statistics of the blocking time
((\ref{xdef}), (\ref{gum}), Figure~\ref{f2}),
and the distribution of the number of spin flips
((\ref{gp}), (\ref{gi}), Figure~\ref{f6}).

\subsubsection*{Acknowledgements}

Interesting discussions with Silvio Franz are gratefully acknowledged.

\end{document}